\documentclass[11pt, conference, compsocconf]{IEEEtran}
\pdfoutput=1
\IEEEoverridecommandlockouts
\usepackage[draft,inline,margin]{fixme}
\usepackage{cite}
\usepackage{amssymb}
\usepackage{amsmath}
\usepackage{graphicx}
\usepackage{tikz}
\usepackage{forloop}

\newcommand{\qed}{\hspace*{\stretch{1}}$\Box$\vspace{4mm}}

\newcommand{\bigo}{\mathcal{O}}
\newcommand{\ignore}[1]{#1}

\title{A New Data Layout For Set Intersection on GPUs$^*$\thanks{A version of this paper appeared in Proceedings of IPDPS 2011.}}

\begin{document}
\author{\IEEEauthorblockN{Rasmus Resen Amossen and Rasmus Pagh}
	\IEEEauthorblockA{IT University of Copenhagen, Denmark\\
	Email: {\tt \{resen,pagh\}@itu.dk}}}

\maketitle


\begin{abstract}
Set intersection is the core in a variety of problems, e.g. frequent itemset mining and sparse boolean matrix multiplication.
It is well-known that large speed gains can, for some computational problems, be obtained by using a graphics processing unit (GPU) as a massively parallel computing device.
However, GPUs require highly regular control flow and memory access patterns, and for this reason previous GPU methods for intersecting sets have used a simple bitmap representation.
This representation requires excessive space on sparse data sets.
In this paper we present a novel data layout, {\sc BatMap}, that is particularly well suited for parallel processing, and is compact even for sparse data.

Frequent itemset mining is one of the most important applications of set intersection.
As a case-study on the potential of {\sc BatMap}s we focus on \emph{frequent pair mining}, which is a core special case of frequent itemset mining.
The main finding is that our method is able to achieve speedups over both Apriori and FP-growth when the number of distinct items is large, and the density of the problem instance is above 1\%.
Previous implementations of frequent itemset mining on GPU have not been able to show speedups over the best single-threaded implementations.
\end{abstract}



\section{Introduction}
Graphics processing units (GPUs) are currently the technology that gives the largest computing power per dollar (measured in floating-point operations per second).
Developing algorithms for GPU computation is challenging, since the architecture imposes many requirements on the way algorithms work, if the potential is to be fully utilized.
In particular, programs need to be structured in identical threads with as little conditional code as possible (i.e., having regular control flow), such that all threads can run the same instruction at the same time.
Also, the memory access pattern of threads that execute together needs to be highly regular to approach the theoretical bandwidth of the GPU memory.
For computation intensive tasks the availability of hundreds of processing units has resulted in large speedups compared to CPU computation (see e.g.~the survey~\cite{GPU-survey}). 
Even for data intensive tasks such as sorting, advantage over CPU computation has been demonstrated (see e.g.~\cite{conf/sigmod/GovindarajuGKM06,1376670,GPU-samplesort}).

Many computational problems depend on being able to perform set intersection efficiently.
For example:
\begin{itemize}
\item in \emph{boolean matrix multiplication} of two matrices, $M$ and $M'$, we want to find all pairs $(i,j)$ for which $\exists k:M_{i,k} M'_{k,j} > 0$, or equivalently for $A_i=\{j|M_{i,j}>0\}$ and $B_j = \{i|M'_{i,j} > 0\}$, the pairs $(i,j)$ for which $A_i\cap B_j \neq \emptyset$
\item in a database context we might ask for a \emph{join-project} of two tables, i.e., a join of two tables followed by a duplicate eliminating projection that projects away the join attribute.
This is equivalent to sparse boolean matrix multiplication \cite{amossenpagh09joinproject}, and thus dependent on efficient set intersection as well
\item \emph{frequent itemset mining} asks, given a set of transactions $T_1,\dots,T_m$, where $T_i\subseteq \{1,\dots,n\}$, to report all sets $S\subset \{1,\dots,n\}$ having \emph{support} at least $s$ in the transactions.
The support of $S$ is defined as the number of transactions that have $S$ as a subset.
The special case where itemsets are limited to size two (where only item \emph{pairs} are found) is also the core problem when larger itemsets are allowed, and frequent itemset mining in general therefore reduces to efficient set intersection
\item all \emph{conjuctive queries} can be thought of as set intersections: given a dataset $D$, and two pre-processed subsets of data, $f,g:D\to \{0,1\}^{|D|}$, the conjuctive query $\{d\in D|f(d) \wedge g(d)\}$ is exactly equivalent to an intersection.
\end{itemize}

In this paper we consider the general problem of intersecting sets.
However, we use frequent itemset mining as a case study throughout the text, as it is one of the most studied problems that can be solved by reduction to multiple set intersections.
We furthermore focus on itemsets of size two (frequent pair mining), since this special case already has many applications (such as finding binary associations) and is highly challenging when there are many frequent items.
At the end of the paper we outline how our approach could be generalized to deal with larger itemsets.

\paragraph{Set representations in frequent itemset mining}
There are two principal ways of representing a set of transactions. 
In the standard {\em horizontal\/} format the transactions are stored one by one (possibly sorted), whereas the {\em vertical\/} format stores, for each item $i$, the set $S_i$ of indices of transactions that contain $i$.
This set is sometimes referred to as the \emph{tidlist} of $i$.
Observe that finding the support of $\{i,j\}$ is simply a matter of computing $|S_i\cap S_j|$.
If $S_i$ and $S_j$ are stored in sorted order it is an easy task to do this in time $\bigo(|S_i|+|S_j|)$.
If the number of distinct items $n$ is large we see that it is easy to parallelize the computation of all support counts: simply distribute the intersections among the processors such that each processor is responsible for support counts involving a small number of items.

For some data sets (especially sparse ones) it may be faster to use a horizontal layout and maintain a data structure that counts the occurrences of all pairs.
Then the time spent on a pair $\{i,j\}$ is proportional to the support of $\{i,j\}$ rather than to the sum of support of $\{i\}$ and $\{j\}$.
However, this approach may use excessive space when there are many pairs of frequent items.
In parallel and distributed settings the high space usage translates into either using an expensive shared memory, or a phase where the support counts from different parts of the transactions are combined. 
In either case, the communication among processes becomes a bottleneck as the number of frequent items grows.

\subsection{This paper}

\paragraph{Theoretical contribution}
We present a new data format for sets, {\sc BatMap}, that is especially well-suited for parallel and pipelined computation.
It is instructive to compare our format to {\em bitmaps}, which have previously been used to store the sets $S_i$, using one bit per transaction~\cite{GPU-apriori}.
To compute the support of $\{i,j\}$ one needs to perform the bit-wise {\tt AND} of the bitmaps encoding $S_i$ and $S_j$, and count the number of 1s.
This task parallelizes very well, as the bitmaps can be split into any desired number of pieces to be processed individually, and there is a low communication overhead in combining the counts.
It is also very friendly to modern pipelined processor architectures, since no conditional code is needed, avoiding the branch mispredictions that have haunted previous frequent itemset mining algorithms using ``vertical data formats'' and set intersections~\cite{fimi-benchmark}.
Finally, since data can be accessed sequentially, bitmaps make optimal use of cache and prefetching.

The {\sc BatMap} maintains these advantages, while being more space-efficient on sparse sets.
The space usage is in fact within a small factor of the information theoretical minimum for representing sets of a given size, which is the largest imaginable compression.
That is, if $S_i$ and $S_j$ are represented using batmaps $B_i$ and $B_j$ we can compute the size of $S_i \cap S_j$ using a word-by-word comparison of $B_i$ and $B_j$.
In contrast to normal compressed representations of sparse bitmaps, the steps of this computation are completely fixed, and parallelize immediately.
The name {\sc BatMap} indicates the similarity to the functionality of a bitmap, and suggests that this is something that Bruce Wayne might use to mine associations between criminals and crimes.

We should mention a limitation of batmaps compared to bitmaps: 
the result of combining two batmaps is not a batmap, so it cannot directly support the intersection of more than two sets. 
Towards the end of the paper we outline possible ways of dealing with this limitation.

\paragraph{Experiments}
In Section~\ref{sec:experiments} we investigate the performance characteristics of our algorithm (on GPU), and CPU implementations of Apriori \cite{apriori} and FP-growth \cite{fp-growth} for varying density and number of distinct items.
We find that our algorithm scales well in the number of distinct items, in terms of both computation time and memory usage.
In addition, the algorithm performs well for dense instances.

The throughput of batmap intersection on GPU is found to be about 5 times larger than when running the algorithm on the 8 CPU cores on our system.
We also perform experiments comparing batmaps on GPU with merging of sorted lists, a standard CPU-based algorithm for computing intersection size.

\subsection{Previous work}

\subsubsection{Set intersection}

The algorithm for intersecting two sorted lists is folklore.
In the literature it has been extended in two main directions.
The first is {\em adaptive\/} intersection procedures, that use fewer comparisons when there are compact witnesses for the intersection, see e.g.~\cite{conf/soda/DemaineLM00}.
In the worst case, and in the average case, these algorithms provide no speedup over the classical algorithm.
Second, for dense sets there has been considerable work on compressed representations, usually referred to as {\em compressed bitmaps}.
The {\em density\/} of a set is its size divided by the size of the universe from which its elements come (e.g., in the case of frequent itemset mining, the density of $S_i$ is $|S_i|/m$).
Previous work on high-performance compressed bitmap formats include Boncz~\cite{Boncz-compress}, BBC~\cite{bbc} and WAH~\cite{WAH}.
These methods all require data to be decoded sequentially, and provide no easy parallelization.

Bille et al.~\cite{conf/isaac/BillePP07} present a compressed bitmap format that is nearly optimal wrt.~the amount of data read to compute set operations.
However, this is mainly a theoretical result that is not likely to perform well in a GPU setting.
Our new vertical data layout can be viewed as a kind of compressed bitmap, with special properties.

\subsubsection{Frequent itemset mining}

To ease the exposition we will assume that we have preprocessed the data set to remove items with support below the threshold we are interested in.
All existing frequent itemset methods do this, in one way or another, so the interesting comparison is for the case where there are only frequent items.

\paragraph{GPU computation}
The previous work most closely related to ours is that of Fang et al.~\cite{GPU-apriori}.
They use (in the PBI-GPU algorithm) a bitmap to store a vertical representation of the data set.
This means that the representation of a data set of $m$ transactions with $n$ distinct items requires $mn$ bits of space.
For a sparse data set with a total of $mb$ items, where $b\ll n$, this can be much more than the $\log\binom{mn}{mb} \approx mb \log(n/b)$ bits needed to represent the data.
Experiments in~\cite{GPU-apriori}, on hardware similar to what we use, show that their GPU/bitmap is more than 1 order of magnitude faster than a tuned implementation of the Apriori algorithm in some cases where the data set is dense (density 49\%).
For a sparse data set (density 0.6\%) there is basically no speedup.
So both from a space usage and a computation time perspective this method does not work well for sparse data sets.
Based on the experimental results on the synthetic dataset T40I10D100K reported in~\cite{GPU-apriori} we can estimate the speed of the underlying set intersections to be around 40 Gbit per second. 
In the case of T40I10D100K, which has a density of 4\%, this means that they can in 1 second intersect sets of total size around $1.6\cdot 10^9$. 
Sets with lower density take proportionally longer per item, and sets with larger density take proportionally less time.

We also note that~\cite{GPU-apriori} did not present experiments showing that a GPU implementation can be faster than FP-growth~\cite{fp-growth} (in fact, in all three experiments reported, FP-growth was considerably faster).


\paragraph{CPU computation}
A lot of work has been devoted to parallel and distributed implementations of frequent pattern mining.
The survey of Zaki~\cite{Zaki:1999:PDA} describes the state-of-the-art as of 1999.
More recent work has focused on multi-core architectures of modern commodity hardware, trying to optimize cache performance and minimize the overhead of access to shared data~\cite{conf/vldb/GhotingBPKNCD05,conf/vldb/LiL07}.
However, GPU parallelism involves many constraints on the structure of the code and memory access pattern that is not addressed in these works.
In particular, our method exploits the massive SIMD parallelism that is available on GPUs, and we find it conceivable that the set representation we describe could lead to other advances in parallel and distributed computation.

\section{BatMaps}\label{sec:cuckoo}

Let $S_i$ denote the set of transactions containing item $i$. 
We wish to preprocess the sets $S_i\subseteq \{1,\dots,m\}$ such that we can quickly compute the intersection sizes $|S_i\cap S_j|$ for all item pairs $\{i,j\}$. 
A standard solution to this problem is to store the sets as sorted lists, which allows an intersection to be computed in time $\bigo(|S_i|+|S_j|)$ by simple merging.
However, the control flow for this intersection procedure is unpredictable, which makes it work poorly on modern architectures, in particular GPUs, since they require highly structured control flow to perform well.

The initial idea is to rely on hashing rather than comparisons. 
If we organize the sets in hash tables (say, using linear probing or perfect hashing) it is indeed fast to determine the common elements of two sets $S_i$, $S_j$ as we simply look up all elements from $S_i$ in $S_j$.
Using perfect hashing (perhaps with vectorization~\cite{conf/cidr/BonczZN05}) the control flow becomes deterministic and predictable.
However, the memory access pattern of hash table lookups remains random and highly irregular.

Our new approach starts with an old idea from parallel and distributed data structures~\cite{SICOMP::StockmeyerV1984,JACM::UpfalW1987,DietzfelbingerMey93}, applied in a novel way.
The idea is to store sets redundantly to enable more efficient parallel/distributed operations. 
More specifically, we consider the case where an element $x$ can only be stored in the memory locations given by $2d-1$ random hash functions (applied to $x$).
By storing an element in $d$ out of the $2d-1$ possible locations, we get that for any two sets both containing $x$ there is at least one position that contains $x$ in both representations.
This means that it suffices to do a data independent element-by-element comparison which parallelizes very well (see top part of Figure~\ref{fig:comparison}).
\ignore{
\begin{figure}
\newcounter{xv}
\begin{center}
	\begin{tikzpicture} 
		\draw[step=0.25cm,color=black] (-3,0) grid (3,0.25);
		\draw[step=0.25cm,color=black] (-3,0.999) grid (3,1.25);
\forloop{xv}{1}{\value{xv} < 25}{{
\draw [<->](-3.125+\value{xv}/4,0.30) -- (-3.125+\value{xv}/4,0.95);
}
}
	\end{tikzpicture}
	
	\bigskip
	
	\begin{tikzpicture} 
			\draw[step=0.25cm,color=black] (-3,0) grid (3,0.25);
			\draw[step=0.25cm,color=black] (-1.5,1.499) grid (1.5,1.75);
	\forloop{xv}{1}{\value{xv} < 13}{{
	\draw [<->](-3.125+\value{xv}/4,0.3) -- (-1.65+\value{xv}/4,1.45);
	}}
	\forloop{xv}{1}{\value{xv} < 13}{{
	\draw [<->](-0.125+\value{xv}/4,0.3) -- (-1.6+\value{xv}/4,1.45);
	}}
	\end{tikzpicture}	
\end{center}
\caption{Computing the common elements in two batmaps is done using pairwise comparisons. For the sake of the illustration we have drawn each batmap as an array. For batmaps of the same size, we simply need to compare elements at the same position (top). For batmaps of different sizes, each entry in the smaller batmap needs to be compared to several entries in the larger batmap (bottom).}\label{fig:comparison}
\end{figure}
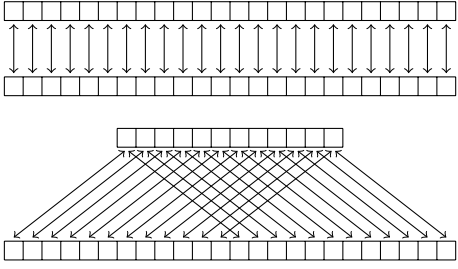
}

\paragraph{Our adaptation} 
We will consider $d=2$ and store each element $x\in S_i$ in two of three hash tables.
For the time being we will simply think of these hash tables as a $3\times r$ array $A^{(i)}$ (section~\ref{sec:implementation} describes the specific layout we use). 
In each hash table $t\in\{1,2,3\}$ there is exactly one position $(t,h^{(i)}_t(x))$ where $x$ may be stored, given by the hash function $h^{(i)}_t$.
There is a probability that the arrangement of values in the hash tables, as illustrated in Figure~\ref{fig:assignment}, is not possible. 
We discuss this probability in Section~\ref{sec:analysis}, and for the sake of the discussion we temporarily assume that the arrangement is always possible.
\begin{figure}
\begin{center}
\begin{tabular}{|p{3mm}|p{3mm}|p{3mm}|p{3mm}|p{3mm}|p{3mm}|p{3mm}|p{3mm}|}
	\hline
	 & & ${\underline x}z$ &  &  &  & ${\underline y}$ & \\
	\hline
	 & & ${\underline z}$ &  &  & ${\underline x}y$ & & \\
	\hline
	 ${\underline y}$ & &  & $x{\underline z}$ &  & & & \\
	\hline
\end{tabular}
\end{center}
\caption{Example of a 2-of-3 assignment for the set $S=\{x,y,z\}$. Each element has one possible position in each of the three hash tables. Two of these (where the element is underlined) are used to store the element.
}\label{fig:assignment}
\end{figure}

It will be important that all sets are stored according to the {\em same} hash functions $h_1$, $h_2$, $h_3$, with range scaled according to the size of the set. 
That is, given hash functions $h_1$, $h_2$, $h_3$, we let $h^{(i)}_t(x) = h_t(x) \text{ mod } r_i$, where $r_i= \bigo(|S_i|)$ is a power of two to be specified later. 
Since we choose ranges that are powers of 2, observe that for $r_{i}<r_{j}$ we have $h_t^{(i)}(x)=h_t^{(j)}(x) \text{ mod } r_{i}$. 
This means that if $x\in S_{j}$ is stored in $(1,p_1)$ and $(2,p_2)$ it suffices to check positions $(1,p_1\mod{r_{i}})$ and $(2,p_2\mod{ r_{i}})$ to determine if $x\in S_{i}$. 
Below, we explain how this principle can be used to efficiently count the number of items in $S_i\cap S_j$.
We will return to the issue of constructing the representation later.

Suppose that $x\in S_i\cap S_j$. 
Then, because we have stored $x$ redundantly in the hash tables there exists at least one $t$ for which $A^{(i)}_t[h_t^{(i)}(x)] = A^{(j)}_t[h_t^{(j)}(x)]$.
For now we assume that $r_i = r_j = r$, which means that $h_t^{(i)}=h_t^{(j)}$ for all $t$.
Now, by making all equality checks of the form ``$A^{(i)}_t[p]==A^{(j)}_t[p]$'', where $t\in\{1,2,3\}$ and $p\in \{0,\dots,r-1\}$, we can identify each element in $S_i\cap S_j$.
These comparisons, illustrated in Figure~\ref{fig:comparison}, parallelize very well.
However, to count the number of elements in the intersection, an additional trick is needed.
We can impose a cyclic order to the three hash tables, such that $h_1$ is followed by $h_2$, $h_2$ is followed by $h_3$, and $h_3$ is followed by $h_1$.
Then for an occurrence of $x$ in a hash table it makes sense to ask whether the other occurrence of $x$ is in the hash table is before or after (it will be in exactly one of these).
We use a single bit per position $p$ in the hash tables to store this information, denoted $b^{(i)}_t[p]$.
Consider a pair of items $\{i,j\}$, and a position $(t,p)$ in their batmaps (assumed to be of the same size).
In order to only count exactly once a transaction $x$ where both items appears, we use the condition $(A^{(i)}_t[p]=A^{(j)}_t[p]) \wedge (b^{(i)}_t[p] \vee b^{(j)}_t[p])$ to determine if the elements in position $p$ are overlapping and should be counted.
See Figure~\ref{fig:assignment2} for an illustration.
\begin{figure}
\begin{center}
\begin{tabular}{|p{3mm}|p{3mm}|p{3mm}|p{3mm}|p{3mm}|p{3mm}|}
	\hline
	 & $x0$ &  &  &  & \\
	\hline
	 & &  &  & $x1$ & \\
	\hline
	 &  & &  & & \\
	\hline
\end{tabular}\\
\vspace{2mm}
\begin{tabular}{|p{3mm}|p{3mm}|p{3mm}|p{3mm}|p{3mm}|p{3mm}|}
	\hline
	 & &  &  &  & \\
	\hline
	 & &  &  & $x0$ & \\
	\hline
	 &  & $x1$ &  & & \\
	\hline
\end{tabular}\\
\vspace{2mm}
\begin{tabular}{|p{3mm}|p{3mm}|p{3mm}|p{3mm}|p{3mm}|p{3mm}|}
	\hline
	 & $x1$ &  &  &  & \\
	\hline
	 & &  &  & & \\
	\hline
	 &  & $x0$ &  & & \\
	\hline
\end{tabular}
\end{center}
\caption{The three possible 2-of-3 assignments with respect to a single element $x$. Along with each occurrence is the bit that tells whether this occurence is before or after the other occurrence in the circular order of rows. When counting the common elements in two data structures we use this information to only count the last occurrence, in case the data structures store an item $x$ in the same two positions. This is accomplished by a logical {\tt OR} of the associated bits.}\label{fig:assignment2}
\end{figure}
It is easy to check that in both the case where an element $x$ is stored in the same two hash tables in both batmaps, and the case where there is only one overlapping occurrence, $x$ is counted exactly once.
We will see later that there will be positions $p$ in each of $A_t^{(i)}$ that contain no element from $S_i$ --- in these positions we simply set $A^{(i)}_t[p]=\bot$ and $b^{(j)}_t[p]=0$ to ensure that no counting is done.
Here $\bot$ is a NULL value that is not in any set $S_i$.

For the general case, since $r_i$ divides $r_j$, each position in $A^{(i)}$ corresponds to $r_j/r_i$ positions in $A^{(j)}$ as explained above.
That is, we can again count the number of elements in $S_i\cap S_j$ by comparing each position in $A^{(j)}$ with a position in $A^{(i)}$ (see Figure~\ref{fig:comparison}).

\paragraph{Compression}
Since our method is based on hashing we can use a compression scheme that stores each item relative to the set of items with the same hash value (see section~\ref{sec:layout} for details). 
This gives a significant space saving for dense sets:
in our implementation each hash table entry uses just 8 bits, including $b^{(i)}_t[p]$, whenever the density of a set is above~$2^{-8}$.

\subsection{Data structure construction}
We employ an insertion procedure that generalizes cuckoo hashing~\cite{cuckoo-jour} (which places elements in 1 of 2 possible positions).
The idea is to push elements around until an element is placed in a vacant position (with content $\bot$).
An insertion of $x$ starts by putting $x$ in $A_1$, kicking out any element that might reside in $A_1[h_1(x)]$, making it {\em nestless}.
In case there is a nestless key, it is inserted in $A_2$ in the same fashion, and so on using the circular order $1,2,3,1,2,3,\dots$.
If the number of element moves exceeds a threshold {\tt MaxLoop} the procedure returns the element that is currently nestless (our analysis below shows that this is a small probability event).
The pseudo code is as follows (where $\leftrightarrow$ is used to denote the swapping of two variable values).

\begin{minipage}{\linewidth}
\begin{tabbing}
  xx\=xx\=xx\=xx\=xx\=\kill
  \+\+\\
  {\bf function} {\sc insert}$(\tau)$\+\\
    {\bf loop} MaxLoop {\bf times} \+\\
    $\tau \leftrightarrow A_1[h_1(\tau)]$\\
    {\bf if} $\tau=\bot$ {\bf then return} $\bot$\\
    $\tau \leftrightarrow A_2[h_2(\tau)]$\\
    {\bf if} $\tau=\bot$ {\bf then return} $\bot$\\
    $\tau \leftrightarrow A_3[h_3(\tau)]$\\
    {\bf if} $\tau=\bot$ {\bf then return} $\bot$\-\\
    {\bf end loop}\\
    {\bf return }{$\tau$}\-\\
  {\bf end}
\end{tabbing}
\end{minipage}

\medskip

Since we need two occurrences of each element $x$, the insert procedure is called twice for each element.
In case one of these insertions fails, we delete any occurrences of $x$ and re-insert the nestless element returned (unless it happens to be identical to $x$).
In the Analysis section below we bound the probability of insertions to fail.
While this probability is low for a single set, failed insertions are likely to occur when handling many sets.
We describe how we handle failed insertions in Section~\ref{sec:prepostprocessing}.

\subsection{Analysis}\label{sec:analysis}

Suppose we have a data structure for a set $S$, with hash functions of range $r$.
We now consider what might happen when we insert an element $x_1$ using the insert procedure. 
Possibly, a single copy of $x_1$ has already been inserted in the hash table. 
All other elements exist in exactly two copies. 
When moving an element it may happen that it is moved to the location of the other copy of that element. 
In this case the other copy is then moved to the third location, which must contain a different element. 
We consider the {\em transcript} of the insertion, which is the sequence of values of the variable $\tau$ from the {\sc insert()} function after each element move upon insertion of $x_1$.

We first look at the possibility that each {\em copy\/} of an element appears only once in this sequence, i.e., that each element appears at most twice. 
Then each prefix of the transcript has the form $x^{d_1}_1,x^{d_2}_2,\dots,x^{d_{k}}_{k}$, where $x_1,\dots,x_{k}$ are distinct and $d_1,\dots,d_{k}\in \{1,2\}$ (number of copies that we move).
Each such sequence appears with probability $r^{1-k}$, since we have a hash collision between $x_i$ and $x_{i+1}$ for $i=1,\dots,k-1$, and each such collision happens independently with probability at most $1/r$. 
Taking the union bound over all choices for $x_2,\dots,x_{k}$ and $d_1,\dots,d_{k}$ we get an upper bound on the probability that a transcript prefix of length $k$ occurs:
$$2^{k} n^{k-1} r^{1-k} = 2\,(2n/r)^{k-1}.$$

The next case to consider is when the transcript involves the same copy of an item more than once (a {\em loop}).
Then it is not hard to realize that the insert procedure will move a prefix of the elements in the transcript back to their original positions, and eventually have $\tau = x_1$ again.
Then $x_1$ is pushed to a new table, and we again have two cases to consider.
\begin{enumerate}
\item The transcript does not again return to an element copy that appeared previously.
Consider a prefix of the transcript of length $k'$. 
Then at least one of the two substrings of the transcript of length $k = \lfloor k'/3\rfloor$ that start with $x_1$ will have no repeated element copies.
We can bound the probability of such a transcript in the same way as above:
$$2^{k} n^{k-1} r^{1-k} = 2 (2n/r)^{k-1} \leq 2 (2n/r)^{k'/3-2}.$$
\item The transcript returns once again to a previously visited element copy (a second loop).
Let $k$ denote the number of distinct elements encountered. 
The number of transcripts starting with $x_1$ is then at most $2^k k^2 n^{k-1}$, where the $k^2$ factor is an upper bound on the number of ways the two loops can be formed. 
There are $k+1$ independent hash collisions for such a transcript, so each has probability $r^{-k-1}$, and by a union bound we see that this is an unlikely event when $r\geq (2+\varepsilon)n$:
$$2^k k^2 n^{k-1} r^{-k-1} = (2n/r)^k k^2 / (nr).$$
\end{enumerate}
Notice that the insertion may fail only in the last case.
Using the assumption that $r\geq (2+\varepsilon)n$ we see that this happens for some $k$ with probability at most 
\begin{align*}
& \sum_{k=1}^n (2n/r)^k k^2 / (nr)\\
& \leq (nr)^{-1} \sum_{k=1}^n k^2 (1+\varepsilon/2)^{-k}\\
& = \bigo((\varepsilon^{3}nr)^{-1}).
\end{align*}
Here, we have bounded the sum by computing the integral wrt.~$k$ from $0$ to $\infty$.

When the insertion succeeds, we see that the probability that it goes on for $k'$ steps or more is bounded by $2 (2n/r)^{k'/3-2}$.
Thus, the expected number of steps is bounded by
\begin{align*}
&\sum_{k'=1}^\infty 2 (2n/r)^{k'/3-2}\\
& \leq \sum_{k'=1}^\infty (1+\varepsilon/2)^{-k'/3+2}\\
& = \bigo(1/\varepsilon).
\end{align*}
Thus, by choosing $\varepsilon > 0$ as a constant, the expected time for performing all insertions is $\bigo(n)$.


\section{Implementation}\label{sec:implementation}

The implementation is split into two parts: code for execution at the GPU, and the pre- and postprocessing on the host system (CPU).

\subsection{Layout of data structures}\label{sec:layout}

Our actual implementation differs a bit from the abstract description in Section~\ref{sec:cuckoo}.
We compress the data so that only 8 bits are used per batmap element, while still being able to handle densities larger than $2^{-8}$.
Define three permutations, $\pi_t:\{1,\dots,m\}\to\{1,\dots,m\}$ for $t\in\{1,2,3\}$, let as earlier $r_i$ denote the domain size of the hash functions for batmap $B_i$, and define the hash functions $h_t^{(i)}$ by
\begin{align*}
h_t^{(i)}(x) & = |B_0|\left\lfloor\frac{\pi_t(x)\text{ mod } r_i}{r_0}\right\rfloor\\
     & \quad + (\pi_t(x)\text{ mod } r_0) + (t-1)r_0.
\end{align*}
The batmap layout induced by these hash functions is illustrated in Figure~\ref{fig:newrowlayout}.
\ignore{
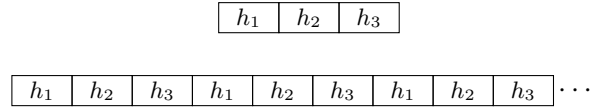
\begin{figure}
\centering
  \begin{tikzpicture}[scale=0.8]
    \footnotesize
    \def\w{1.0}
    \foreach \i in {1,...,3} {
      \draw(\w*\i,0) rectangle (\w*\i+\w,0.5);
      \draw (\w*\i+\w/2,.25) node{$h_\i$};
    }
  \end{tikzpicture}\\
\vspace{5mm}
  \begin{tikzpicture}[scale=0.8]
    \footnotesize
    \def\w{1.0}
    \foreach \j in {0,...,2} {
      \foreach \i in {1,...,3} {
        \def\o{3*\w*\j}
        \draw(\o+\w*\i,0) rectangle (\o+\w*\i+\w,0.5);
        \draw (\o+\w*\i+\w/2,.25) node{$h_\i$};
      }
    }
    \normalsize
    \draw(10.4*\w,.25) node{$\cdots$};
  \end{tikzpicture}
\caption{Organization of the three hash functions for $B_0$ (top) and $B_i$ (bottom) where $|B_0|=3r_0$.
Each $h_t$ above represents $r_0$ batmap elements covered by that hash function.}\label{fig:newrowlayout}
\end{figure}
}
An important observation is now, that instead of storing element $x$ at position $h_t^{(i)}(x)$ we could just as well store $\pi_t(x)$ at that position---the result of the element-wise comparisons between two batmaps would be the same.
Next, by definition of $h_t^{(i)}$ the position of $\pi_t(x)$ (the stored representation of $x$) in a batmap uniquely identifies the least significant bits in $\pi_t(x)$, so explicitly storing these can be considered superfluous.
Therefore, instead of storing $x$ we will only store the 7 most significant bits of $\pi_t(x)$.
That is, $\pi_t(x)$ can now be deduced from the position and the 7 bits stored in that position.
Furthermore, we use 1 additional bit per batmap element to store the indicator bit $b^{(i)}_t[p]$ described in Section~\ref{sec:cuckoo}, and organize the bits so the indicator bit is the most significant of the 8 bits.
This compression gives us 4 elements per 32-bit integer.

To get an idea of the efficiency of this compression scheme, assume that we have to shift $s$ bits to the right in order to move the 7 most significant bits down to the least significant bits.
Then $\log(m+1)-s\leq 7$, and consequently $2^s \geq (m+1)/128$.
Also, as each element's position in a batmap should uniquely identify the least significant $s$ bits of $h_t^{(i)}$ all hash domains must be at least of size $r_i\geq 2^s$ for this compression to work.
If we compare to the uncompressed case with hash domain sizes of $2\cdot2^{\lceil\log(|S_i|)\rceil} \approx 2|S_i|$, we only obtain an actual compression (space reduction) when the input is sufficiently dense, i.e. where the set size is satisfying $2|S_i| \geq 2^s$, or equivalently $|S_i| \geq (m+1)/256$.

In the GPU, the actual comparisons are done in chunks of 32-bit integers (4 batmap elements at a time) in a way that completely avoids conditional statements:
let $x$ and $y$ denote two 32-bit integers, and for convenience, let the 7 least significant bits in each 8-bit block be referred to as the \emph{element bits} as they refer to a batmap element.
If $\oplus$ denotes a logical {\tt XOR} and ``$(\cdots)_{16}$'' means hexadecimal notation then
$$p=((x \oplus y) \vee ({\tt 80808080})_{16}) - ({\tt 01010101})_{16}$$
gives a 0 (not 1) in the indicator bits iff the corresponding element bits of $x$ and $y$ are equal.
To negate these bits, and only count a match if one of the corresponding indicator bits is set, define
$$p'=(p\oplus ({\tt ffffffff})_{16}) \wedge ((x \vee y) \wedge ({\tt 80808080})_{16}).$$
We then account for $((p'\gg 7) + (p'\gg 15) + (p'\gg 23) + (p'\gg 31)) \wedge 7$ matches among the $2\times 4$ elements represented by $x$ and $y$.
Here, $\gg$ denotes the shift operator as usual.

\subsection{Our adaption of the GPU execution model}\label{sec:out_gpu_adaption}
The execution model in GPUs and OpenCL can roughly be outlined as follows:
a \emph{kernel} is a set of instructions to be evalutated on a set of cores in a multiprocessor, and a thread running such a kernel is in OpenCL referred to as a \emph{work item}.
These work items can be organized in a one, two or three dimensional grid of size $W_1\times W_2 \times W_3$, also referred to as a \emph{work group}, and each running kernel instance can retrieve its coordinate (\emph{local index}) in this grid.
Also, we define the \emph{global} data size as a multiplum of the work group size $G_1W_1 \times G_2W_2 \times G_3W_3$.
When executing the kernel, work groups are generated by iterating over the global size, i.e. a total of $G_1G_2G_3$ work groups are formed.
As with the local index, each kernel instance can retrieve its work groups' current global coordinate (\emph{global index}) in this iteration process.
As an example, consider a kernel that processes a two-dimensional $3200\times 3200$ pixels image in chunks of $16\times16$ tiles.
This would correspond to a work group size of $16\times 16$ threads, a global data size of $3200\times 3200$, and consequently $200\cdot200=4000$ work group positions in the global data.

OpenCL operates with multiple memory spaces, but here we will only refer to two of these:
the most plentiful memory space, \emph{global} memory, is the only memory space accessible from the host device (the CPU), and it has the largest latency among all the memory spaces.
The low-latency \emph{shared} memory resides closer to each compute unit, it is relatively small (e.g. around 16 kb), and is shared among all the threads in a work group.
One of the most important considerations when implementing efficient algorithms for execution at GPUs is coalescing global memory accesses, and we achieve this by following best practice as described in \cite{nvidia}.
In short, global memory access by threads of a \emph{half warp} (16 threads) are coalesced by the device in as few as one transaction when certain access requirements are met, e.g. if the 16 threads access a 64 bytes aligned segment, corresponding to 16 32-bit integers.

We adapt the GPU execution model to the ideas described in Section~\ref{sec:cuckoo} and \ref{sec:layout} in the following way:
a list containing all $n$ batmaps is transferred once to the device, and we then define the global size to be $n\times n$, and the work groups to be of size $16\times 16$.
Consequently, a total of $n^2$ batmap comparisons will be made, in chunks of size 16.
The thread with local index $(l_i,l_j)$ and global index $(g_i, g_j)$ will now handle the comparisson of batmap $B_{16g_i + l_i}$ and $B_{16g_j + l_j}$ in turns of 16 integers (holding 64 batmap elements):
each of the 256 threads in the work group first copies two single items from the input, which resides in global memory, into two small $16\times 16$ integer arrays in shared memory.
Each row in these small arrays correspond to a 16 integer wide slice of batmap $B_{16g_i}$ to $B_{16g_i+15}$, and $B_{16g_j}$ to $B_{16g_j+15}$, respectively.
Because of coalescing, this copying is very efficient.
Second, after synchronising the threads with a \emph{memory barrier}, the 16-item wide batmap slices are now compared as described above, and the process is repeated with another copying from global to shared memory.
This continues until all slices of the relevant batmaps have been compared.

\subsection{Pre- and post processing}\label{sec:prepostprocessing}
As the batmap comparisons are performed in the GPU in quantums of 2 times 16 consecutive batmaps the computation time of each such 16-block will be determined by the longest of these batmaps.
Therefore, as a first step, we sort the batmaps by increasing width (corresponding to sorting the sets $S_i$ by size), resulting in a strongly reduced computation time for the subresults for narrow batmaps.
That is, after sorting we have $|B_i| \leq |B_j|$ for $i<j$.

Many graphics devices have a few-second hard limit on the execution time when the device is also used to support the display.
Therefore, we break the GPU calculation into smaller parts of size $k\times k$ where $k$, in our experiments, typically had a value of 2048.
Let $Z_{p,q}$ be a matrix holding the subresults for batmaps $B_{pk}$ to $B_{pk+k-1}$ and $B_{qk}$ to $B_{qk+k-1}$.
The division into smaller sub problems now has the convenient side effect that we, due to symmetry, only need to compute $Z_{p,q}$ for $p\leq q$, thereby cutting almost half of the GPU computation time, from $n^2$ to around $\binom{n}{2}$.

\paragraph{Failed insertions}
As there is a positive probability that some of the cuckoo insertions will fail due to collisions with previously inserted elements we need to handle these failed insertions separately.
Let $F_b$ be the set of items $i$ for which insertion of value $b$ in batmap $B_i$ failed, and let $A_b$ denote \emph{all} items in input associated with $b$.
For all transactions $b$, we construct the pairs $(\min(a,c), \max(a, c))$ for which $a\in F_b$ and $c \in A_b$, and store each pair in a set $M_{p,q}$ where $(p,q) = (\lfloor\min(a,c)/k\rfloor, \lfloor\max(a, c)/k\rfloor)$.
Whenever a subresult $Z_{p,q}$ is returned from GPU we extend it with the pairs found in $M_{p,q}$ before reporting the number of pairs found.
(For $p=q$, only the upper triangle of $Z_{p,q}$ is reported because of symmetry.)


\section{Experiments}\label{sec:experiments}

\paragraph{Hardware setup}
All experiments were run on a MacPro with two Intel Xeon 5462, 2.8 GHz, 4-core CPUs and 6 GB RAM (bus speed 1.6 GHz), running Mac OS X 10.6.
The machine had a GeForce GTX 285 graphics card with 1 GB RAM and 30 1.4 GHz cores having 8 computation units each.
We observe that the two Xeon chips (combined) and the GPU have a similar complexity, with a total of 1.6 and 1.4 billion transistors, respectively\footnote{Manufacturer's specification.}.
However, the price of the 2 CPUs is significantly higher than that of the GPU (the factor is around 5 based on Intel's initial price for Xeon 5462, but this ratio has likely decreased somewhat).
A specified indicator of the maximal energy consumption (TDP) is $2\times 80$ W for the Xeon CPUs, and $204$ W for the GPU, so the energy consumption at full utilization is likely to be similar.




\subsection{Frequent pair mining}
In this section we report on experiments on frequent pair mining. Readers who are primarily interested in the raw performance of set intersections may skip ahead to the paragraph {\em Throughput computation}.

We have implemented frequent pair mining with batmaps in Python, using the PyOpenCL interface to OpenCL.
Even though it is would be possible to parallelize individual set intersection computations, we have chosen to focus on the case where the number of items is large, such that it suffices to run the different intersections in parallel.
The output of our algorithm is the support of every pair of items.

We will compare our algorithm with Apriori \cite{apriori, conf/fimi/Borgelt03,conf/fimi/Borgelt04} and FP-growth \cite{fp-growth, Borgelt-FP-growth}---both implemented by Christian Borgelt.
Some experiments on Eclat \cite{eclat} were also performed but it was significantly slower than the other three implementations and has therefore been left out of the graphs.
Even though other implementations have been reported to be faster in some cases (e.g.~\cite{conf/fimi/Racz04,conf/fimi/UnoKA04}), we found that the implementations available in the FIMI repository did not compile with recent versions of {\tt gcc}.
Thus, we have settled for Borgelt's implementations, that are generally regarded as state-of-the-art, as witnessed by a total of 35 citations in 2008-2010.
Each test run had a hard limit of 1800 CPU seconds before it was cancelled.

The first set of experiments illustrate the behavior of the three algorithms when keeping the instance size constant and varying either the number of distinct items or the item density.
An instance was generated by, for each transaction, including each of the $n$ distinct items with probability $p$, and continue adding transactions until the desired total instance size was reached.

\begin{figure}
\includegraphics[width=.45\textwidth]{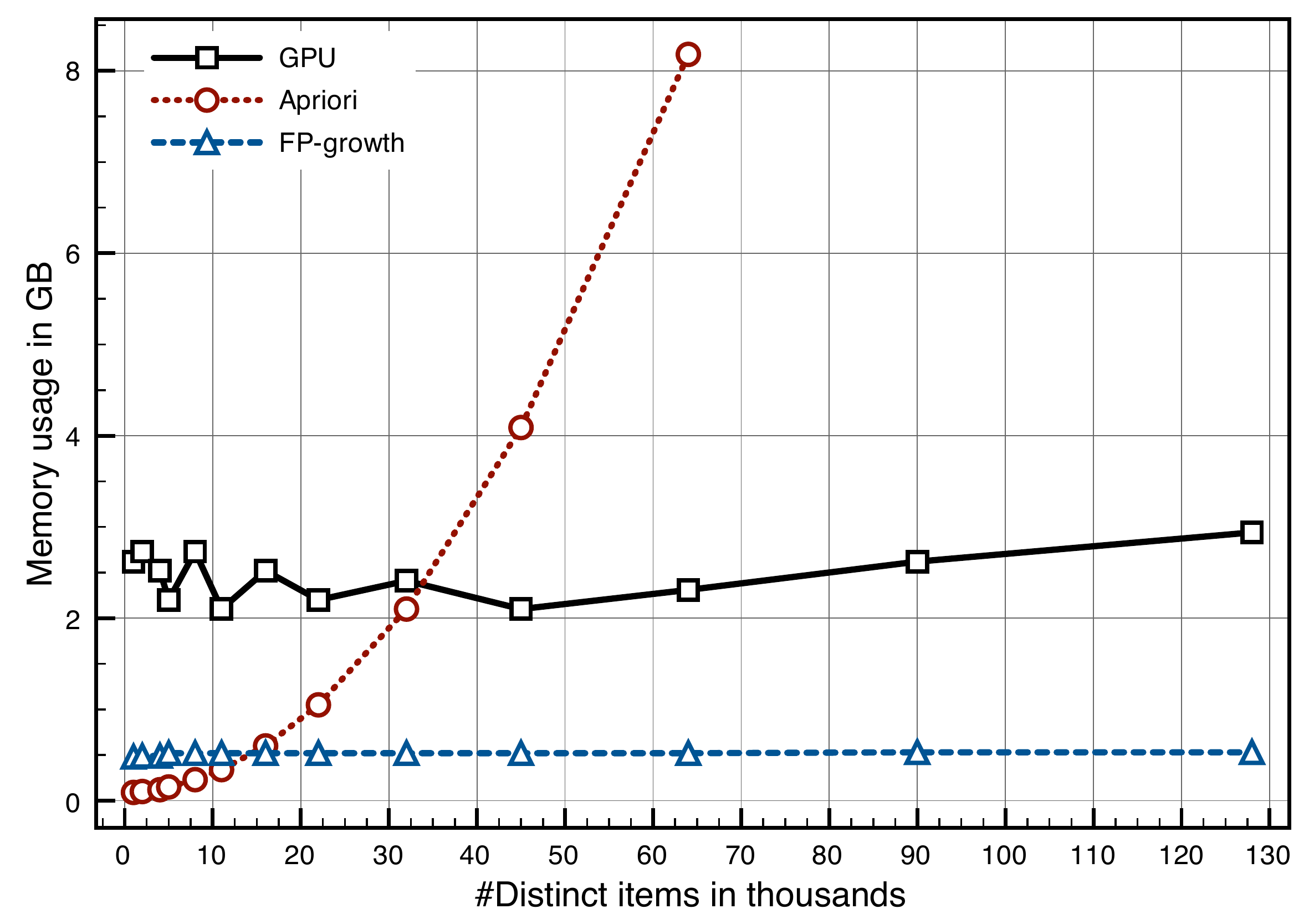}
\caption{Memory usage for varying number of distinct items $n$, while holding the instance size at a constant 10 million items with an item density of 5\%. 
Apriori scales poorly with $n$.}\label{fig:nitems_mem}
\end{figure}

Figure~\ref{fig:nitems_mem} depicts the memory usage for the three algorithms for varying number of distinct items $n$.
The space usage of the GPU implementation comes from the preprocessing, which is done on the CPU.
We did not attempt to optimize the space usage of our preprocessing procedure, so it is likely that significant savings could be obtained by a space-aware implementation.
From the plot we se that while both FP-growth and the GPU implementation scale well with~$n$, Apriori has quadratic memory usage and exceeds the 6 GB RAM for less than 64,000 items.

\begin{figure}
\includegraphics[width=.45\textwidth]{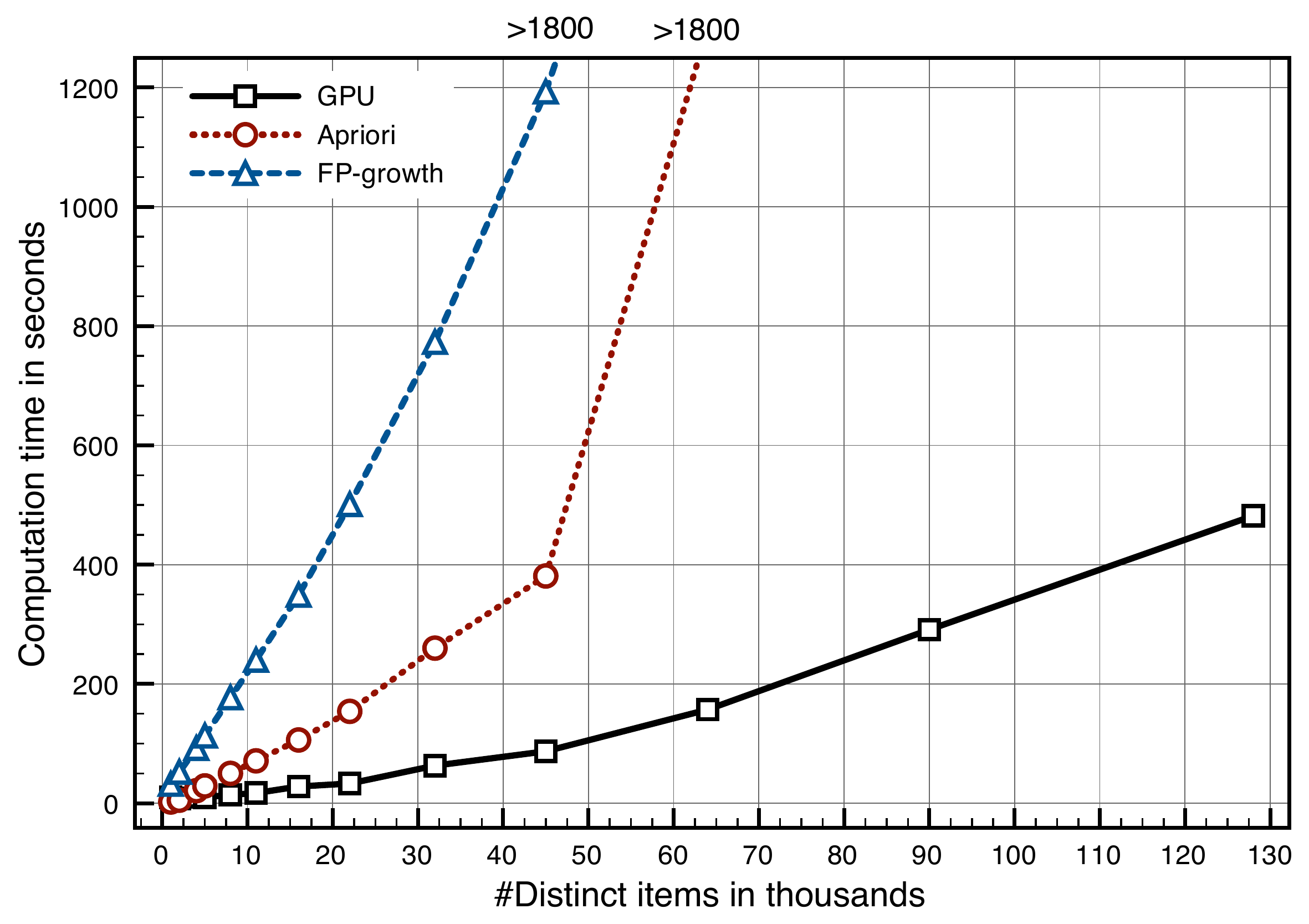}
\caption{Computation times on pure pair generation for varying number of distinct items, while holding the instance size at a constant 10 million items with an item density of 5\%. Both Apriori and FP-growth exceeds their time limit on 1800 seconds when solving the $n=64,000$ instance. In comparison, the GPU implementation scales well in $n$.}\label{fig:nitems_core}
\end{figure}

Figure~\ref{fig:nitems_core} compares the pure pair generation times for varying number of distinct items, but keeping the data size fixed.
This is the part of all three methods that has super-linear complexity, so focusing on this allows us to see the asymptotic behavior more clearly.
Not surprisingly $n=64,000$ is an upper bound on what can be run with Apriori within the time limit, due to memory trashing.
As expected, FP-growth exhibits linear growth in time usage as the number of items increases.
The GPU algorithm has space and time usage that grows linearly with the number of distinct items, but is more than 1 order of magnitude faster than FP-growth (on a single core).

\begin{figure}
\includegraphics[width=.45\textwidth]{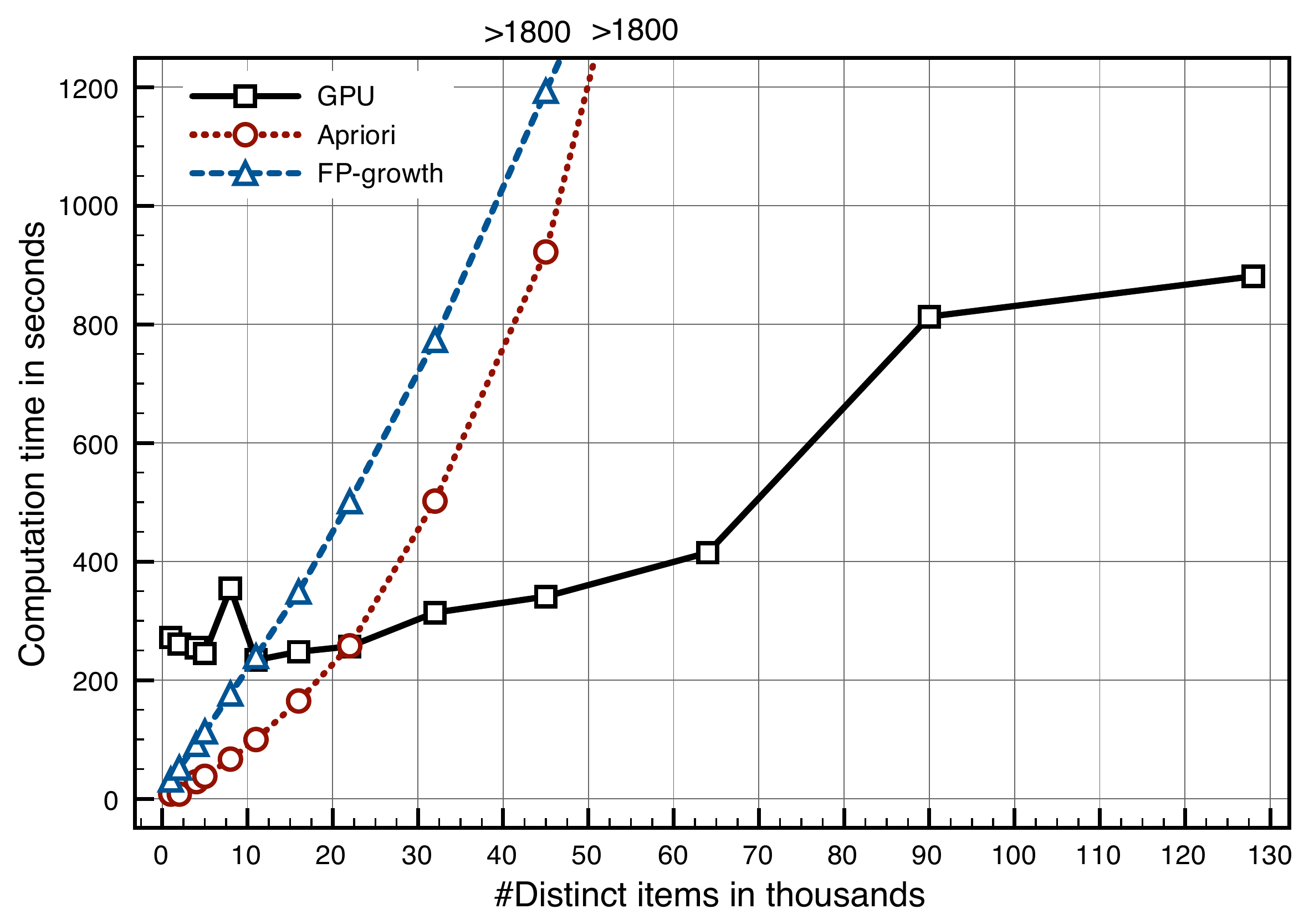}
\caption{Total computation times, including pre- and postprocessing for varying number of distinct items, while holding the instance size at a constant 10 million items with an item density of 5\%.
The preprocessing time for the GPU implementation is high, but scales well in $n$.}\label{fig:nitems_total}
\end{figure}

Figure~\ref{fig:nitems_total} shows the total execution times including pre- and postprocessing.
Our implementation suffers from high preprocessing times, partly due to our choice of Python (which is interpreted) as language. Still, our implementation outperforms Apriori and FP-growth for large $n$.
According to a popular benchmark~\cite{pythonbench}, Python executes between 2 and 106 times slower than GNU \texttt{C++} with a median of 49.
We therefore believe that an optimized implementation of the preprocessing in {\tt C} would achieve at least 1 order of magnitude speedup compared to our simple Python implementation.

\begin{figure}
\includegraphics[width=.45\textwidth]{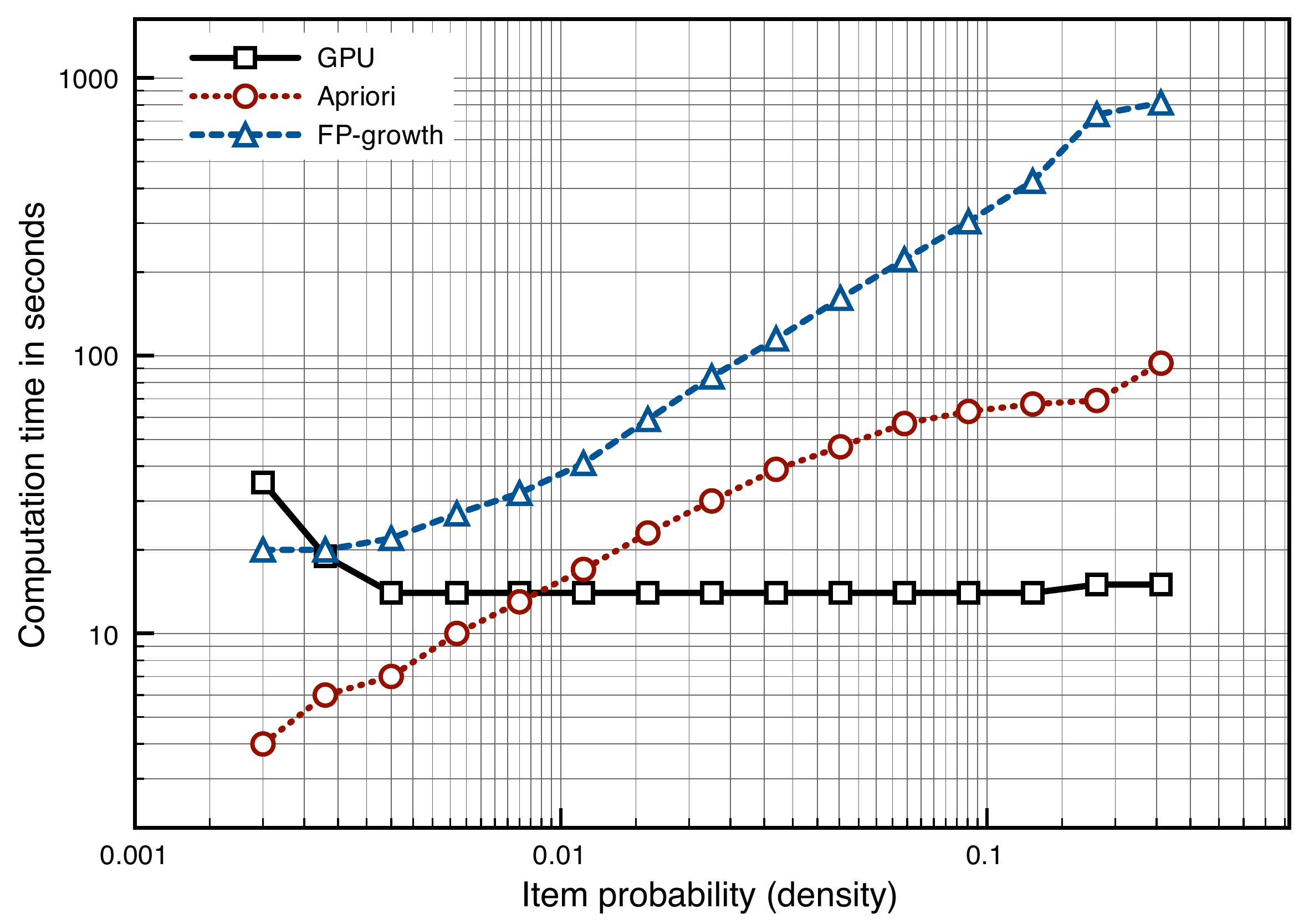}
\caption{Computation times on pure pair generation for varying item density, while holding the instance size and number of distinct items constant at 10 million and 8000, respectively.}\label{fig:prob_core}
\end{figure}

We tested the behavior of the algorithms for varying item densities, and the results can be seen in Figure~\ref{fig:prob_core}.
While both Apriori and FP-growth have difficulties handling dense instances, our GPU implementation uses time almost independent of density.
It can be noticed that for low densities the GPU time actually increases.
This is due to the lower bound on space requirement for our compression scheme as described in Section~\ref{sec:layout}.

\begin{figure}
\includegraphics[width=.45\textwidth]{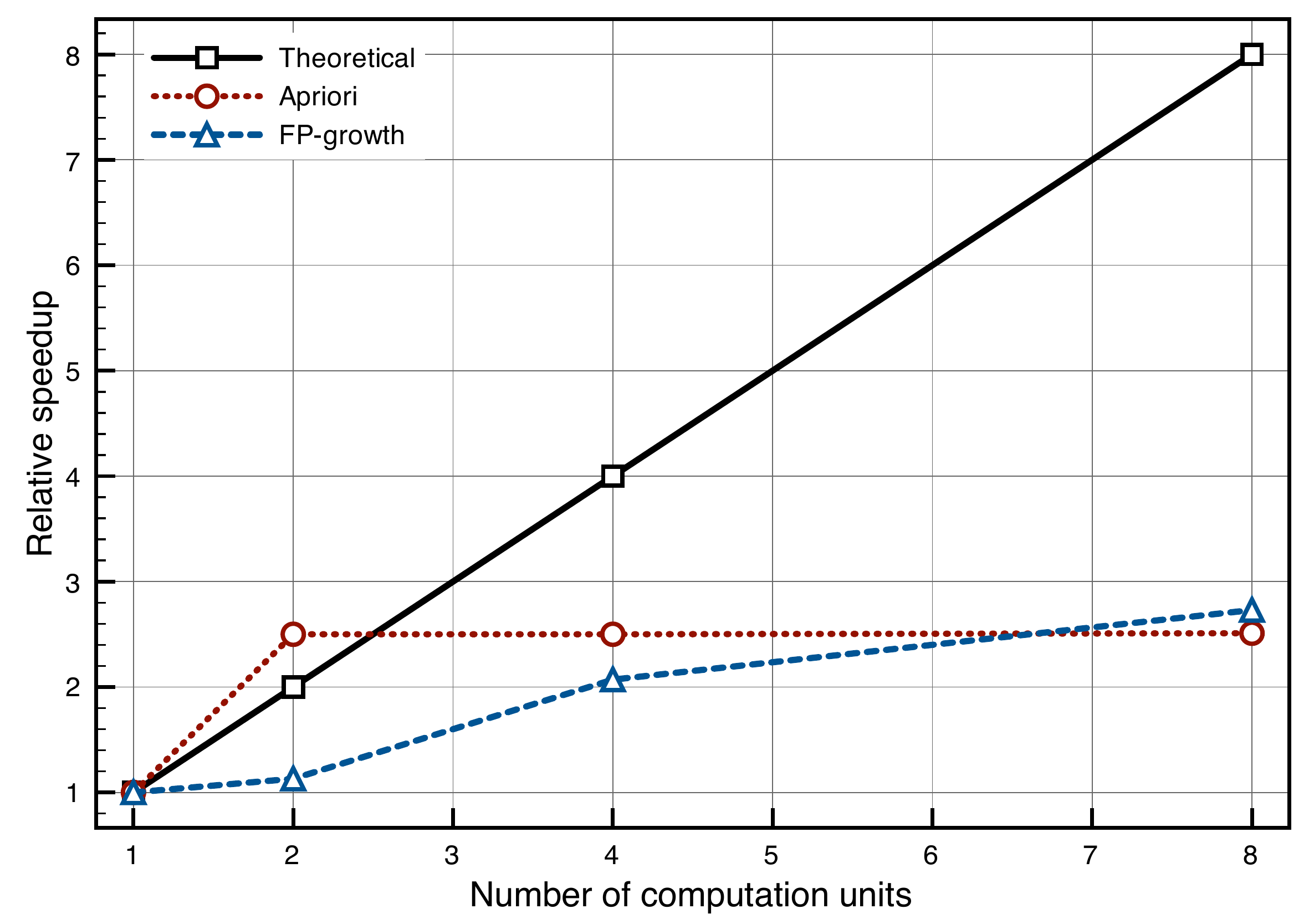}
\caption{The relative speed-up vs. the number of computation cores. The theoretical speed-up is linear, but neither the implementation of Apriori nor FP-growth were benefitting noticably from more than four cores.}\label{fig:ncores_speedup}
\end{figure}

In Figure~\ref{fig:ncores_speedup} we try to illustrate how Apriori and FP-growth might scale to a larger number of computation cores.
Our experiments was based on an instance of size 10 million items, 4000 distinct items, and a density of 5\%.
In a test simulating parallel execution on $i$ cores, we split the original instance into $i$ smaller instances of identical size.
We compare the maximum execution times of test runs for $i \in\{1,2,4,8\}$.
As seen in the figure, none of the algorithms benefit noticeably from more than four cores.
This is consistent with previous work which also finds that Apriori scales poorly on many processors~ \cite{conf/sera/YeC06}.

\begin{figure}
\includegraphics[width=.45\textwidth]{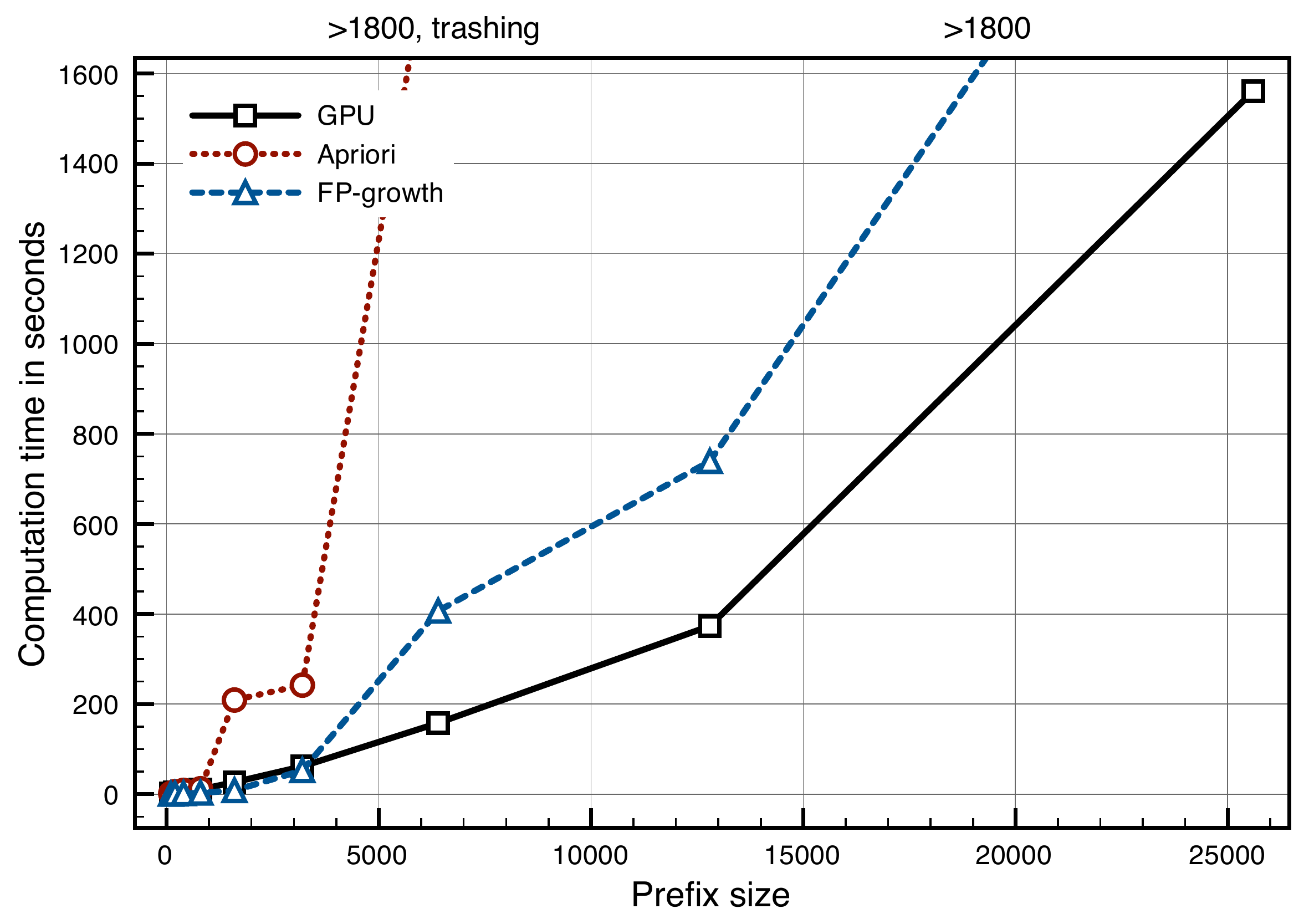}
\caption{Computation time for pure pair generation for increasing prefix sizes of the WebDocs instance.
The number of distinct items increases rapidly which explains why the computation time for Apriori explodes for small prefixes.
None of the algorithms could solve a prefix of size 51,200 within the 1800 seconds time limit, and the memory usage of helper data structures for the GPU implementation exceeded the 6 GB RAM available.
}\label{fig:webdocs_core}
\end{figure}

The last experiment, seen in Figure~\ref{fig:webdocs_core}, compares the algorithm performances on a ``real-life'' data set, WebDocs, which associates web documents and words.
The data set was taken from the Frequent Itemset Mining Dataset Repository\footnote{\tt http://fimi.cs.helsinki.fi/data/}.
As WebDocs is an enormous instance we run several tests on prefixes of varying size.
The number of distinct items in this instance increases rapidly so all three algorithms are challenged.
As seen, Apriori exceeds the time limit first due to memory trashing.
The GPU algorithm solves the largest instance: a 25.600 line prefix.

\medskip

\paragraph{Throughput computation}
The number of items processed by the GPU for a pair mining run can be estimated as follows.
Consider the experiment with $n=4000$ distinct items, a total instance size of $10^7$, and $p=5\%$.
Sets in this instance have average size $10^7/4000 = 2500$, which means that each batmap is  $3\cdot 2^{\lceil\log(2\cdot 2500)\rceil} = 3\cdot 2^{13}$ bytes wide.
Thus the combined input size to all set intersections is $4000^2 \cdot 3\cdot 2^{13}$ bytes.
The experiment used $10.87$ seconds on the GPU and thus we processed 36.2 Gbyte per second.
The memory bandwidth on the GPU, however, is around $159$ Gbyte per second so we are a factor of over 4 from the theoretical maximum memory throughput.

To get an idea of the performance on GPU relative to the performance of an equivalent implementation on CPU, we performed the following experiment:
two arrays of 5,000,000 32 bit integers were created, element-wise comparison using the counting technique described in Section~\ref{sec:layout} was performed 300 times, and the total execution time was measured.
The size of the arrays was chosen to measure the performance on non-cache-resident data.
The implementation was written in {\tt C}, and compiled with {\tt gcc} with optimization level {\tt O3}.

\begin{figure}
\includegraphics[width=.45\textwidth]{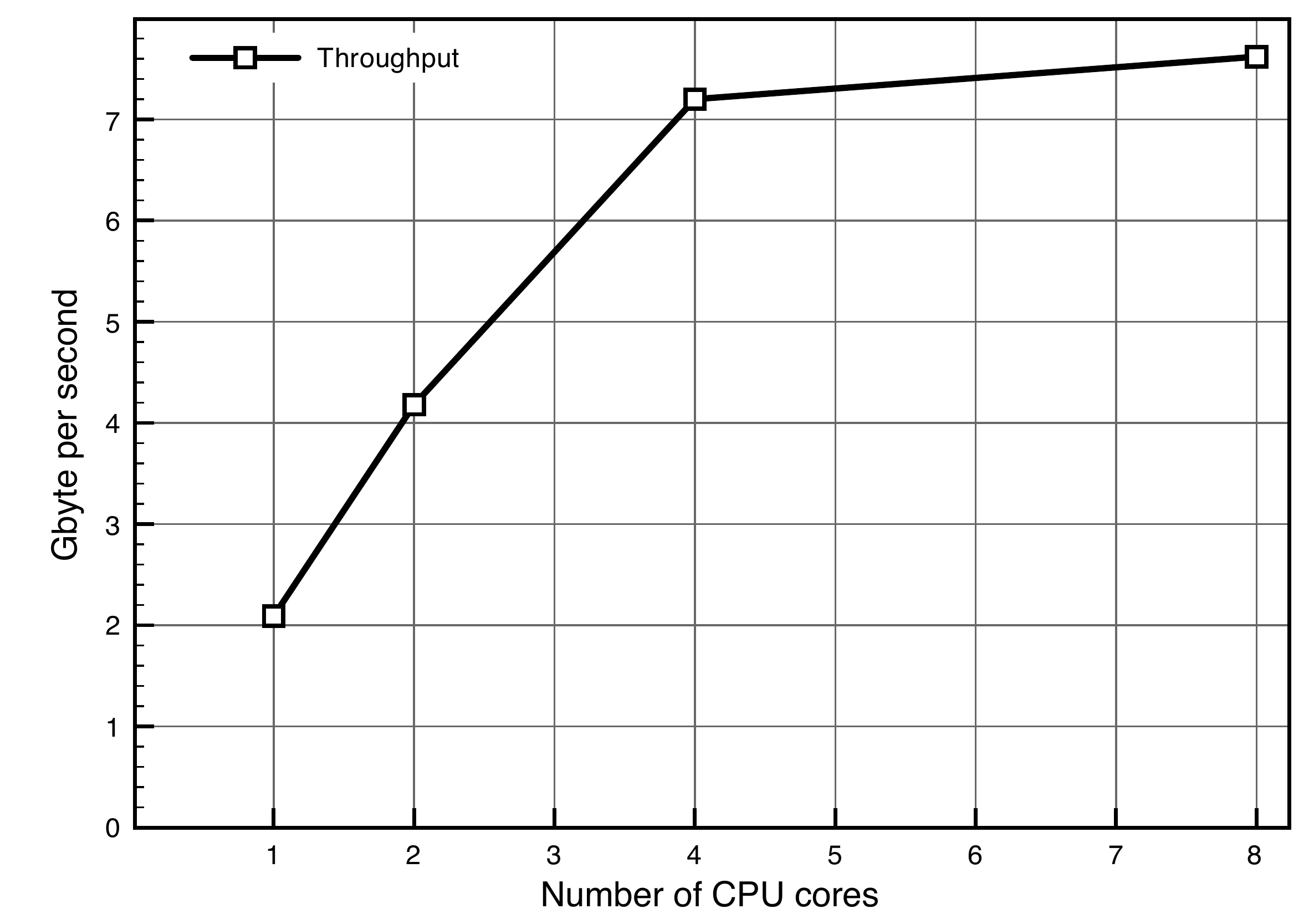}
\caption{Memory throughput of CPU comparison of batmaps (size 20 Mbyte). The CPU encounters a memory bottleneck when using 4 cores, and the throughput never exceeds 7.6 Gbyte per second which is almost a factor 5 slower than the GPU.}\label{fig:cpu_comparison_speed}
\end{figure}

Figure~\ref{fig:cpu_comparison_speed} shows the average processing speed using 1, 2, 4, and 8 simultaneous CPU cores.
Section~\ref{sec:out_gpu_adaption} described how the GPU implementation divides the complete problem in $16\times16$ tiles, and for each such tile, copies elements from global to shared memory.
In the CPU implementation we ignore the cost of these memory operations.
Still, the processing speed of the CPU never exceeds 7.6 Gbyte per second.
This is almost a factor 5 slower than the 36.2 Gbyte per second obtained on the GPU.



\subsection{Comparison with merging}
A widely used representation of sets, that allows efficient computation of intersections, is sorted lists. 
A simple for-loop can be used to report all common elements, by scanning both lists. 
Even though this algorithm is extremely simple, it runs slowly on modern CPUs due to branch mispredictions.

To compare batmaps on GPU with CPU implementations based on merging, we first compute the number of set elements processed per second in the experiment reported above.
The total input size (in terms of number of set elements) to all set intersections is $4000^2 \cdot 2500 = 40\cdot 10^9$.
Thus, we processed $3.68\cdot 10^9$ elements per second, which is typical for intersections of this size.
Due to rounding of the size of hash tables, batmaps of the same size would be able to accommodate up to 63\% more elements, which would give a maximal processing speed of $6\cdot 10^9$ elements per second.
On the other hand, if the rounding works against us, the processing speed would be only half of this.

We performed an experiment in which we counted the number of identical elements in two sorted arrays of $2^{24}$ integers (32 bits each), repeated 100 times.
The implementation was written in {\tt C}, and compiled with {\tt gcc} with optimization level {\tt O3}.
Doing one such run took 14.89 seconds (on one core), which means that $2.25\cdot 10^8$ elements are handled per second. This is 13--26 times slower than the processing speed on the GPU.

To compare against a parallel implementation, we did 8 simultaneous runs (using 8 cores), which took 15.66 seconds.
Since the time did not grow noticeably, we conclude that the computation does not (yet) have a memory bottleneck.
The number of set elements processed per second using 8 cores is $1.71\cdot 10^9$, or 29--57\% of the throughput of the GPU batmap computed above.
This means that performance is noticeably poorer on the CPUs than on the far less expensive GPU.




\section{Conclusion}

We have shown that a GPU allows set intersection and frequent pair mining that extends to much larger number of items than previous algorithms.
Further, we believe that our approach may be pushed further with careful tuning, as we are still far from using the full memory bandwidth of the GPU.
Our techniques may open up for new applications of e.g.~association mining where there are tens of thousands of variables (e.g.~genetic data).

One problem we leave open is to achieve similar results for intersections of more than two sets.
There are two ways in which our work could possibly be extended: one is to use a generalization of batmaps that store items in $d$ out of $d+1$ places.
This would ensure that itemsets of size up to $d$ would have at least one position witnessing their intersection.
Another is to use batmaps to count, for each item in $S_{i_1}$, how many times this item appears in $S_{i_2}, S_{i_3}, \dots$.
At the end one would need to sum up the counts for the two occurrences of each item to determine if the item appeared in all sets.

\medskip

\section*{Acknowledgement.} 
We would like to thank Anna Pagh for taking part in showing theoretical results on our generalization of the cuckoo hashing insertion procedure, and Kumar Lav for his participation in initial experiments with the method described in this paper. 
Also, thanks to the anonymous reviewers for numerous useful suggestions.
This work was supported in part by a grant from the Danish National Research Foundation for the project ``Scalable Query Evaluation in Relational Database Systems''.

\bibliographystyle{abbrv}


\end{document}